\documentclass{amsart}

\usepackage{mathrsfs, amssymb}
\usepackage{graphicx}
\usepackage{float}

\title{Bifurcation in optimal retirement} 
\author{Bushra Shehnam Ashraf}
\address{CANNEX Financial Exchanges Limited \\ bushra.sashraf@gmail.com}
\author{Thomas S. Salisbury}
\address{Dept. of Mathematics and Statistics, York University \\ salt@yorku.ca}
\thanks{Salisbury's work is supported in part by a grant from NSERC. The authors wish to thank Huaxiong Huang and Moshe Milevsky for their help and suggestions during the course of this work.}

\begin{document}
\pagestyle{plain}

\begin{abstract}
We study optimal consumption and retirement using a Cobb-Douglas utility and a simple model in which an interesting bifurcation arises. With high wealth, individuals plan to retire. With low wealth they plan to never retire. At a critical level of initial wealth they may choose to defer this decision, leading to a continuum of wealth trajectories with identical utilities. 
\end{abstract}

\maketitle


\section{Introduction} \label{Section:DMintro}
We are interested in understanding the kind of utility preferences that drive retirement behaviours. In particular, we will consider a utility of consumption that privileges post-retirement consumption over pre-retirement consumption, and will optimize retirement in that context. The current paper carries this out in the setting of a deterministic hazard rate, and forms a portion of the Ph.D. thesis \cite{Ashraf} of the first author. A subsequent paper will address the same question, but in the context of a stochastic hazard rate. 

Consider an individual of age $x$ at time $t=0$, with a stochastic remaining lifetime $\varsigma$ governed by a deterministic hazard rate $\lambda_t$. They are interested in determining the optimal time to retire while maximizing their expected lifetime utility. Currently, the individual is employed and earning a labour income of $\$ 1$ annually until retirement. They have no bequest motives and post-retirement consumption will be funded purely by savings (i.e., there is no exogenous pension income stream). The individual is willing to invest only in risk-free assets earning an instantaneous return $r$. 

Our goal will be to optimize the time of voluntary retirement $t^*\le T$, where $T<\infty$ is the maximum possible lifetime (here taken to correspond to age $110$). A particularly interesting consequence of using our simple model is that we will find that there is a clear bifurcation in behaviour, with low initial wealth prompting a decision never to retire ($t^*=T$), and high initial wealth leading to reasonably prompt retirement. In between, there will be a critical value of initial wealth at which there will be a continuum of possible retirement ages, all yielding identical utilities. 

This lies in contrast to a subsequent paper (in preparation), in which allowing the hazard rate to be stochastic leads to significantly more complicated possibilities.

\section{Earlier work} \label{Section:DMliterature}

Bodie, Merton and Samuelson \cite{BMS1992} considered the retirement decision problem as an optimal stopping problem. Kula \cite{Kula2003} treated the option to retire as an investment process where we collect retirement wealth and once this accumulated wealth reaches a certain threshold, we retire. Koo, Koo, Shin \cite{KKS2013} considered a voluntary retirement problem with a focus on optimal investment, consumption and leisure under a Cobb-Douglas utility framework. They used dynamic programming strategies to derive a closed form solution for the value function and optimal strategies for consumption, leisure, investment and retirement. 

Farhi and Panageas \cite{FP2007} used an optimal stopping approach to solve for optimal consumption and portfolio choice problem to make the decision about retirement time while adjusting the labor supply. They wrote ``The ability to time one's retirement introduces an option-type character to the optimal retirement decision. This option is most relevant for individuals with a high likelihood of early retirement, that is individual with high wealth levels". 

Francesco and Sergio \cite{FS2021} 
used a martingale approach to solve for optimal consumption, labor supply and portfolio choice to find the optimal time to retire. They used a constant force of mortality and concluded that if a person enters labor market at age 25, then the optimal retirement age should be between 50 to 65 with an average of 55 years. 

\section{Problem formulation} \label{Section:DMsetup}

Our analysis allows a general deterministic hazard rate  $\lambda_t$, but our numerical experiments will use the usual Gompertz law of mortality, that is, $\lambda_t=\frac{1}{b}e^{(\frac{x+t - m}{b})}$. Here $m$ is the modal value of life in years and $b$ is the dispersion parameter in years \cite{MAM2006}. The probability of the individual surviving beyond time $t$ is given by
\begin{equation} 
Pr[\varsigma>t]={_tp_x} =e^{-\int_0^t  \lambda_q dq} 
\label{1}
\end{equation}

Let $c_t$ and $l_t$ represent the consumption rate and leisure rate processes at time $t$, and let  $w_t$ be the individual's wealth at time $t$. Let $V(w,l)$ be the value function defining the objective of the individual. Given the initial wealth level $w=w_0$, the individual wishes to maximize the discounted utility of leisure and consumption while determining an optimal time for voluntary retirement from the labour force. That is: 
\begin{equation} 
V(w,l)=\max _{c_t,{t^*}} \textsl{E} \left[ \int_0^T \! e^{-\rho s}u(l_s,c_s)1_{\{s\leq\varsigma\}}ds \bigg| w_0 =w, l_0=l \right] 
\label{2}
\end{equation}
where $\rho$ is a subjective discount rate and $u$ is the individual's utility function. The per period leisure process $l_t$ \cite{FP2007} is assumed to be a piece-wise constant function. The individual is endowed with a constant $\overline{l}$ units of leisure for retirement.
\begin{equation} 
l_t = \left\{ \begin{array} {ll} 
             l_{1} \text{, for $t$ pre-retirement}\\
             \overline{l} \text{, for $t$ post-retirement.} 
             \end{array} \right.
\label{3}
\end{equation}																
where, $\overline{l} > 1$ and without any loss of generality $l_1$ is normalized to be equal to $1$. In this work, leisure is not denominated in time (as is often the case in the economics literature). Instead it is simply a parameter that dictates how much enjoyment is there in consuming post-retirement as compared to pre-retirement. The utility of individual will be described by a
\emph{Cobb-Douglas} utility function \cite{FP2007}:
\begin{equation} 
u(c,l) = \frac{1}{\alpha}  \frac{{(c^{\alpha}l^{1-\alpha})}^{1-\gamma ^*}} {1-\gamma ^*} 
\end{equation}
Here $\alpha$ (with $0<\alpha<1$)  measures the weight of the consumption contribution to the individuals utility per period. $\gamma ^*$ (with ${\gamma ^*}\neq1$ and $\gamma^* >0$) is the relative risk aversion coefficient of the individual for two goods; consumption $c$ and leisure $l$. For $\gamma^* >1$, $c$ and $l$ are substitutes for each other while for $\gamma^* <1$ they are complements. 

Let $\gamma = 1-\alpha (1-\gamma ^*)$, so utility takes the form
\begin{equation} 
u(c,l) = l^{(1-\alpha)(1-\gamma ^* )}  \frac{c^{1-\gamma}} {1-\gamma} 
\label{4}
\end{equation}
and for any time $0\leq t\leq T$,
\begin{equation} 
u(c_t , l_t)=\left\{ \begin{array} {ll} 
                    {u_{1}(c_t)\text{=}\frac{c_t ^{1-\gamma}} {1-\gamma}},& \text{ pre-retirement}\\
                    {\overline{u}(c_t)\text{=}\overline{l}^{(1-\alpha)(1-\gamma ^* )} \frac{c_t ^{1-\gamma}} {1-\gamma}},& \text{ post- retirement} 
									   \end{array} \right.
\label{5}
\end{equation}
Define
\begin{equation}
V(t,w,l)=\max_{{c_t},{t^*}} \textsl{E} \left[ \int_t^T \! e^{-\rho(s-t)} u(l_s,c_s) 1_{\{s\leq\varsigma\}}ds\bigg|w_t =w,l_t = l \right]
\end{equation}		
Wealth evolves deterministically, so the optimal controls will be deterministic, and using Fubini's theorem
\begin{equation}
V(t,w,l)=\max_{{c},{t^*}} \int_t^T \! e^{-\rho(s-t)}u(l_s,c_s) {_sp_x}ds\text{  s.t.  }w_t =w, l_t = l
\label{6}
\end{equation}	
which takes one of two forms, which we denote																			
\begin{equation} 
V(t,w,l) \equiv \left\{ \begin{array} {ll} 
                {\overline{V}(t,w);l=\overline{l}},& \text{ post-retirement}\\
                {V_{1}(t,w);l=1},& \text{ pre-retirement.} 
								 \end{array} \right.
\label{7}
\end{equation}

The utility from using a consumption stream $c_s$ for $0\leq s\leq t$ and then using optimal behaviour after $t$ is given by
\begin{equation}
Z_t= \int_0^t \! e^{-\rho s}u(l_s,{c_s})_sp_x ds + e^{-\rho t}{} _tp_x V(t,w_t,l_t)
\label{8i}
\end{equation}

As usual, optimality means that $Z_t$ is constant with optimal behaviour and decreasing in general. We'll break this optimization problem into two inter-linked problems: the Pre-retirement case $(t<t^*)$; and the Post-retirement case $(t \geq t^*)$.
\bigskip
\subsection{Pre-retirement} \label{ssec:DMpre}
Consider the pre-retirement case. We expect there to be an optimal retirement wealth level $\bar w_t$ such that with wealth $w\ge \bar w_t$, an individual at time $t$ will immediately retire, making $V_1(t,w)=\overline{V}(t,w)$. Wealth dynamics are that 
$dw_t = (1+r w_t - c_t )dt$, and $w_t$ is constrained to be $\ge 0$ (no borrowing). Differentiating \eqref{8i} gives that in the continuation region $w<\bar w_t$,
\begin{align}
dZ_t=u_1(c_t)e^{-\rho t}& {_tp_x} dt - (\rho +\lambda_t)e^{-\rho t} {_tp_x} V_1(t,w_t)\notag\\
&+e^{-\rho t} {_tp_x}\{{V_1}_t(t,w_t)dt+{V_1}_w(t,w_t)(1+r w_t - c_t)dt\}
\label{9}
\end{align}
Since $Z$ represents maximal utility, we must have $dZ_t\leq0$ and $dZ_t=0$ for the optimal choice of $c_t$ and $t^*$. 
Factoring out $e^{-\rho t} {_tp_x}$, \eqref{9} gives
\begin{equation}
\sup_{c_t} \{u_1(c_t)-{c_t} {{V_1}_w}\}- (\rho + \lambda_t) V_1 + {V_1}_t + {V_1}_w (1+r w_t ) = 0 
\label{11}
\end{equation} 
for $w<\bar w_t$, with smooth pasting at the free boundary $w=\bar w_t$. Optimizing over $c_t$ in the usual way yields optimal consumption $c_t^*=({V_1}_w)^{-\tilde{\gamma}}$ and a PDE (Hamilton Jacobi equation) for $V_1 (t,w); w<\bar w_t$
\begin{equation}
 {V_1}_t -(\rho + \lambda_t)V_1 +(1+r w){V_1}_w-\frac{{{V_1}_w}^{1-\tilde{\gamma}}}{1-\tilde{\gamma}}=0
\label{14}
\end{equation}

\subsection{Post-retirement}  \label{ssec:DMpost}
Once retirement occurs, we have utility $u(c_t , l_t) = \overline{u}(c_t)$ and value function is $V(t, w_t , l_t ) =\overline{V}(t,w_t)$. Therefore the lifetime utility from \eqref{8i} takes the form
\begin{equation} 
\int_0^t \! e^{-\rho s} \overline{u}({c_s})_sp_x ds + e^{-\rho t}{} _tp_x \overline{V}(t,w_t) 
\end{equation} 
Arguing as before leads to an optimal consumption of
\begin{equation}
{c_t}^*={\overline{l}}^{\tilde{\gamma}(1-\alpha)(1-\gamma*)} {\overline{V}_w}^{-\tilde{\gamma}}
\label{19}
\end{equation}
and the PDE (HJB equation)
\begin{equation}
 {\overline{V}}_t -(\rho + \lambda_t)\overline{V} +r w {\overline{V}}_w - \frac{ {\overline{l}}^{\tilde{\gamma}(1-\alpha)(1-\gamma*)} } {1-\tilde{\gamma}} {\overline{V}_w}^{(1-\tilde{\gamma})} =0
\label{20}
\end{equation}

Due to the absence of labour or pension income, a natural scaling relation for $\overline{V}(t,w)$ exists, namely
\begin{equation}
\overline{V}(t,kw)=k^{1-\gamma}\overline{V}(t,w)\nonumber
\end{equation}
from which we conclude that the value function takes the form
\begin{equation}
\overline{V}(t,w)=F(t)\frac{{w}^{1-\gamma}}{1-\gamma}.
\label{21}
\end{equation}
Set $B\equiv \overline{l} ^{\tilde{\gamma}(1-\alpha)(1-\gamma *)}$. After some manipulation, \eqref{20} becomes
\begin{equation*}
F'+r(1-\gamma)F+B\gamma F^{1-\tilde{\gamma}}-(\rho+\lambda_{t}) F = 0
\end{equation*}
Defining $f(t)=F(t)^{\tilde{\gamma}}$ this reduces to the ODE
\begin{equation}
\frac{df}{dt}+{\tilde{\gamma}}[r(1-\gamma) -(\rho + \lambda_t )]f +B =0.
\label{23}
\end{equation}
Because $\overline{V}(T,w)=0$, we have a terminal condition $F(T)=f(T)=0$. We may also compute that $c^*_t=\frac{Bw}{f(t)}$.

One could prove a verification theorem, showing that a smooth solution of our HJB equations will indeed represent the solution to our optimization problem. Instead of doing this we will focus on the behaviour or solutions to our equations, obtained numerically.

\subsection{Numerical Scheme}  \label{Section:DMnumerical}

We use the Matlab routine \emph{ODE 45} to solve the ODE \eqref{23} and obtain the post-retirement value function $\overline{V}(t,w)$.
For the pre-retirement partial differential equation \eqref{14} we use an upwind explicit finite-difference scheme. In order to capture the behaviour of the value function for small values of wealth $0\leq w <1$, a log-transform of wealth variable is applied first. For each backwards time step, we solve \eqref{14} and then compare with $\overline{V}(t,w)$ to decide whether retirement is optimal, and to find $V_1(t,w)$ and then $c^*_t$ and the free boundary $\bar w_t$.

$T=120$ is a common choice for the maximum age at death, but we used $T=110$ for the sake of numerical stability at low wealth levels.

For our numerical results the parameter values used are $r=2.5\%, \rho=r, \alpha=0.5, \overline{l}=6.49, b=9.44, m=88.82$ and $\gamma=2$, unless mentioned otherwise. We used the grid sizes of $dt=0.0005, dw=0.01, dy=0.01$ and cut off wealth at $w=30$. The computations took around 48 hours at this fine grid size. Qualitative as well as quantitative results stayed the same if the grid sizes changed to $dt=0.001, dw=0.02,dy=0.02$. 

\subsection{Calibrating $\overline{l}$} \label{ssec:DMcalibration}
Most parameters used have natural and well understood choices. The exception is $\overline{l}$, for which we required calibration. We searched for an $\overline{l}$ such that with initial wealth $w_0=1$ at age 30, we would see a retirement age in the range 55--65, and a wealth at retirement of 7 to 12 times annual labour income. We drew the latter numbers from an article \emph{How much do I need to save for retirement?}  \cite{Fidelity} at Fidelity.com. The recommendations there are that an individual retiring at age 67 should target a wealth of 10 at retirement, and 7 by age 55. 

Finding a suitable $\overline{l}$ proved surprisingly tricky. Because our free boundary value problem calls for an explicit scheme, we require a very small time step, and have a stability issue at small wealth values (even with the log transform). We were however able to realize reasonable agreement with our target, for $\overline{l}$ in the range from 6.25 to 6.50, which led to our choice of $\overline{l}=6.49$;

\section{Wealth Dynamics} \label{sec:DMwealth}

With $c^*_t$ and $\bar w_t$ now in hand, we may walk forward from $t=0$ and follow the evolution of wealth in time from various choices of initial wealth $w_0$, using the wealth dynamics
\begin{equation*}
\frac{dw}{dt}=
\begin{cases}1+rw(t) -c(t), &t<t^*\\
rw(t)-c(t), & t\geq t^*
\end{cases}
\end{equation*}

In the following figures, recall that the labour income of the individual is set to 1 unit annually, so we are using a wealth scale with units of yearly labour income. That means $w=5$ would represent $5$ times the annual labour income. 

\begin{figure}[H]
	\centerline{\includegraphics[width=0.9\textwidth]{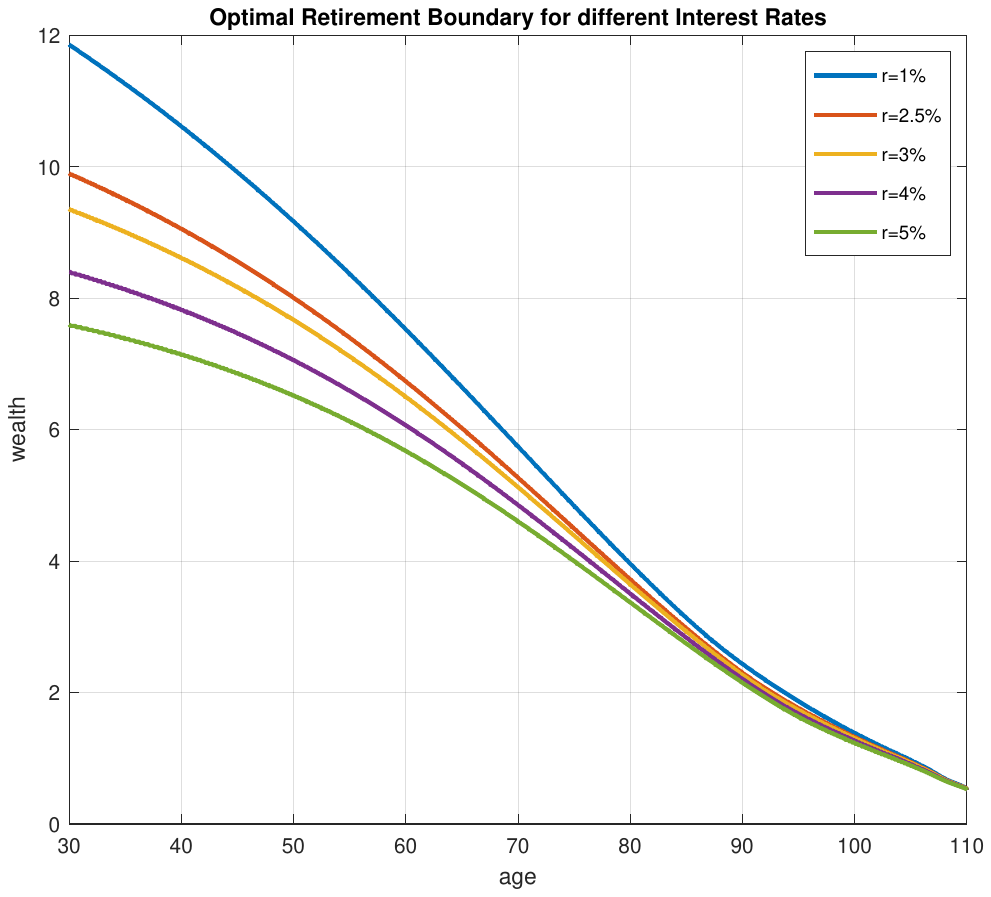}}
	\vspace{-3cm}
	\caption{Optimal retirement age and wealth for different values of annual risk free rate.\newline Assuming $\gamma=2, \rho=2.5\%$, $\alpha=0.5, \overline{l}=6.49, b=9.44, m=88.82, \tau=110$}
        \label{fig:3}        
\end{figure}

Figure \ref{fig:3} plots the optimal retirement boundary $\bar w_t$ as a function of age $x+t$ for various choices of the interest rate $r$. For any $r$, the portion above the corresponding boundary consists of $(t,w)$ for which immediate retirement is optimal. For example, a 30 year old with wealth $9.89$ times labour income will optimally retire immediately, if $r=2.5\%$. 

The curve $\bar w_t$ decreases over time, as older retirees require smaller nest eggs.  as the individual pushes retirement forward in time.  With increased interest rates the retirement boundary shift downwards. This is because the need to accumulate more wealth to sustain post-retirement consumption decreases, since in the post-retirement phase the retirement savings will earn higher interest rates.
\begin{figure}[H]
	\vspace{-4cm}
	\includegraphics[width=0.9\textwidth]{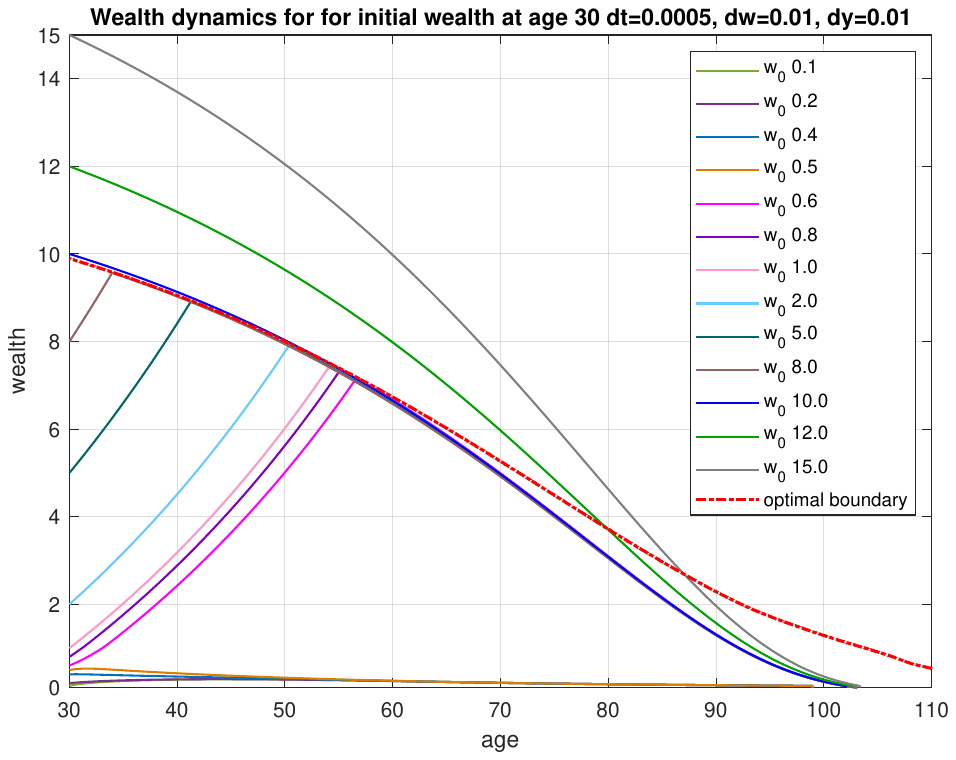}
	\vspace{-4cm}
	\caption{Wealth dynamics for different wealth levels at age 30.\newline Assuming $r=2.5\%, \rho=r$, $\gamma=2, \alpha=0.5, \overline{l}=6.49, b=9.44, m=88.82, \tau=110$ }
	\label{fig:4}
\end{figure}

Figure \ref{fig:4} plots the wealth dynamics of individuals aged 30 who are sitting at different levels of their current savings. As long as the individual is not retired and their wealth-age combination is below the optimal retirement boundary, their wealth follows the pre-retirement dynamics which takes into consideration the labour income. Once they retire, their wealth evolves as per the post-retirement dynamics. If they are above the optimal boundary at age 30, they will not enter the labour force ever. They have enough money saved to spend all their life enjoying the post-retirement life. In this figure, there is a unique strategy followed for each initial wealth plotted.

Note that by age $100$, individuals have mostly exhausted their savings, but some remains. If we had allowed an exogenous pension in the model, we would expect wealth to have actually hit zero by this point. Note also the discontinuity that shows up for low wealth levels, with no retirement and a slow depletion of wealth. In this model, evidently a small change in initial wealth can switch the individual's strategy between saving for retirement, or opting to never retire. If we had incorporated exogenous pension income, this bifurcation would likely not be apparent.

\begin{figure}[H]
	\vspace{-1cm}
	\includegraphics[width=\textwidth]{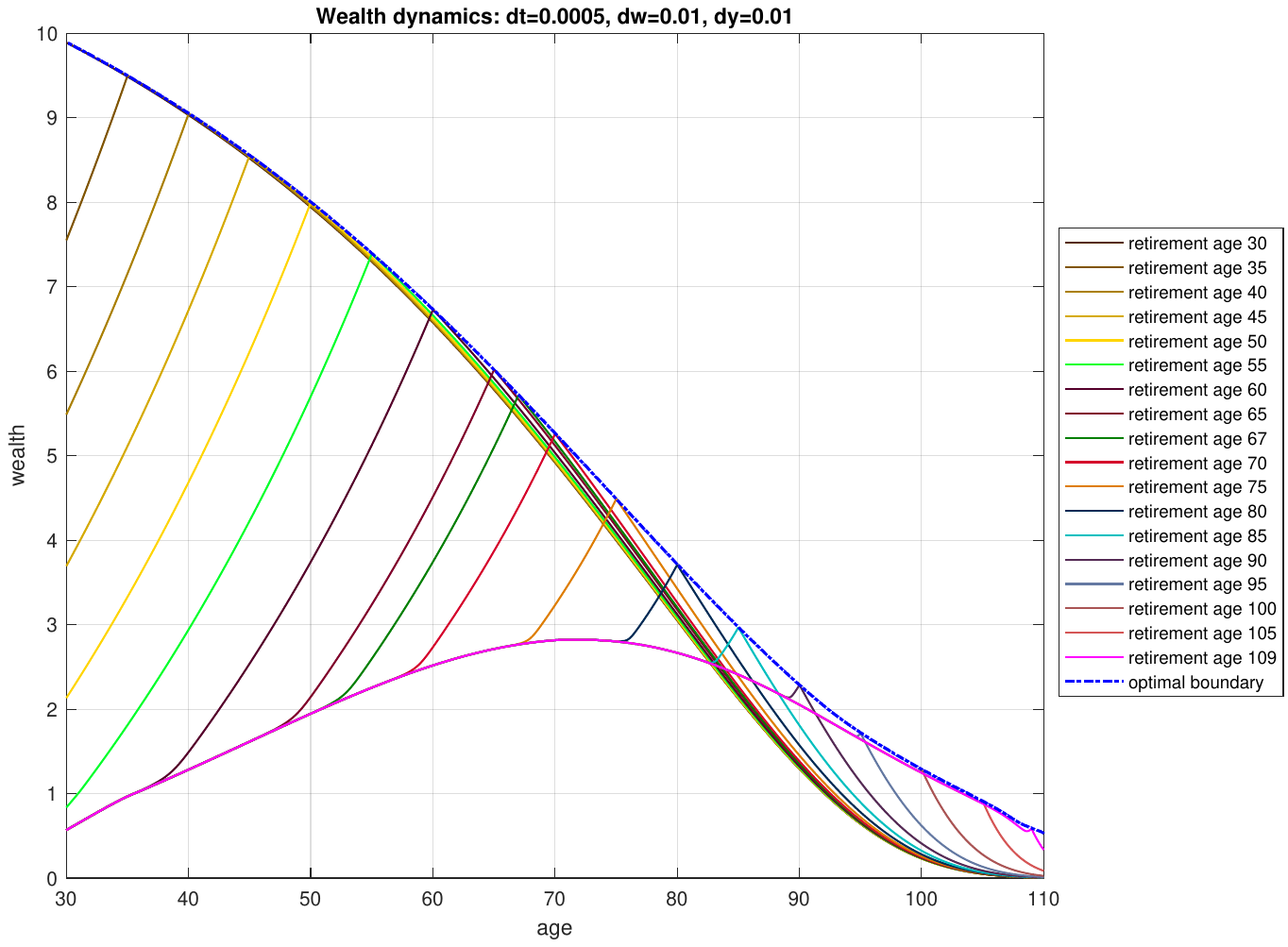}
	\vspace{-1cm}
	\caption{Wealth dynamics for different retirement ages.\newline Assuming $r=2.5\%, \rho=r$, $\gamma=2, \alpha=0.5, \overline{l}=6.49, b=9.44, m=88.82, \tau=110$}
	\label{fig:6}
\end{figure}

We explore this discontinuity in Figure \ref{fig:6}. Here there is a critical initial wealth level $\tilde{w}$ for which multiple strategies yield the same optimal utility. The individual consumes to follow a certain non-uniqueness (or uncommitted) curve that starts at $\tilde w$, without committing to retirement. At any time, they can move up off the curve an infinitesimal amount, which then commits them to a wealth dynamic that leads to retirement. A similar figure could be drawn corresponding to moving down off the curve an infinitesimal amount, and committing never to retire. 

In fact, Figure \ref{fig:6} was drawn by arbitrarily picking certain retirement ages $t$, and following the wealth dynamics forward and backward from $(t,\bar w_t)$. What is observed is that the backwards trajectories actually coalesce. Whereas the forward curves are all scalar multiples of each other, because of the scaling property of the post-retirement value function. 

We see that the time-wealth state space actually consists of three regions. Above the blue optimal boundary, optimal behaviour is to retire right away. Between the blue curve and the magenta curve, optimal behaviour is to save with the intention of retiring. Below the magenta uncommitted curve, optimal behaviour is to plan consumption around never retiring. And for many states $(t,w)$, the only way to reach them is to start with initial wealth $\tilde w$ and to follow the uncommitted curve for a while. 

While our model is a simple one, it has the benefit of providing a clear explanation for the existence of states for which a small perturbation can produce a dramatic change in optimal behaviour. That sensitivity should also be present in many more complex models, without such a clear rationale, so it is useful to understand its origin in this context.

\section{Conclusions}
This paper considers a life cycle model of a consumer with a deterministic (Gompertz) mortality. The consumer is looking to retire at an optimal age and wealth while consuming optimally all their life. We consider the Cobb-Douglas utility framework, in which utility pre- and post-retirement differ by a constant factor (for a given level of consumption). We seek to understand whether basing decisions purely on lifetime consumption with that utility can reproduce realistic behaviour. We indeed find parameter values for which this is the case, though they lie at the limits of stability for our numerical scheme .

The wealth dynamics of the individual exhibit interesting behaviors. If the initial wealth is small, the individual will plan to remain in the labour force all their life. With significant initial savings, the individual will either retire or plan for retirement. But if the initial wealth takes a special value $\tilde{w}$, there are multiple wealth paths that will yield the same optimal utility, involving remaining uncommitted for a time, and then choosing whether or not to target retirement. 

\bibliography{bibintro}{}
\bibliographystyle{plain}

\end{document}